\documentclass[11pt]{article}
\usepackage{geometry}                
\usepackage{graphicx}
\usepackage{amssymb}
\usepackage{epstopdf}
\usepackage{amssymb}
\usepackage{xspace}
\usepackage[binary-units]{siunitx}
\usepackage{lineno}
\usepackage{titlesec}
\usepackage{microtype}      
\titleformat{\paragraph}
{\normalfont\normalsize\bfseries}{\theparagraph}{1em}{}
\titlespacing*{\paragraph}
{0pt}{3.25ex plus 1ex minus .2ex}{1.5ex plus .2ex}
\title{Impact of the HL-LHC detector upgrades on the physics program of the ATLAS and CMS experiments 
\\
}
\author{
Susanne~Kuehn~(CERN) on behalf of the ATLAS and CMS Collaborations
 \\

}
\date{\small{Talk presented at the International Workshop on Future Linear Colliders (LCWS2021), 15-18 March 2021. C21-03-15.1.}\footnote{Copyright 2021 CERN for the benefit of the ATLAS and CMS Collaborations. Reproduction of this article or parts of it is allowed as specified in the CC-BY-4.0 license.}
 }                                          
\begin{document}

\maketitle

\abstract{}  
A wealth of physics results have already been obtained from the LHC, due to the excellent performance of the collider and its experiments. Even more results are expected to be achievable in the phase of the high-luminosity LHC (HL-LHC). It is foreseen to deliver a ten times higher LHC design luminosity resulting in about 4000\,$\mathrm{fb}^{-1}$ within ten years of operation. The upgrade of the LHC is driven by the prospect to observe and measure rare processes. High particle production rates and radiation doses result in a challenging environment for the collider experiments. The ATLAS and CMS experiments are foreseeing to upgrade or even replace several detector components to cope with this environment. In this report an overview of the detector upgrades and their impact on the physics program of the experiments will be given. 

\section{Introduction}
The LHC~\cite{lhc} and its experiments are performing very well with a wealth of physics results. In about six years from now the phase of the high-luminosity LHC (HL-LHC)~\cite{hllhc} is foreseen to start. By a ten times higher LHC design luminosity and luminosity levelling the delivery of about 4000\,$\mathrm{fb}^{-1}$ is envisaged within ten years. The prospect to observe and measure rare processes is driving these
developments. Severe radiation doses and high particle production rates at the HL-LHC result in a challenging environment for the collider experiments. An upgrade or even replacement of several detector components of the ATLAS~\cite{atlas} and CMS~\cite{cms} experiments is anticipated to cope with these circumstances. The aim is for the experiments to have a similar performance as in the Run-2 and Run-3 of the LHC.
In this report several of the detector upgrades and their impact on the physics program of the experiments will be given. 
It starts with an overview of the physics program and motivation for the detector upgrades. Details about the upgrade of several sub-detectors will then be discussed and their impact on physics performance objects highlighted. The report closes with prospects for selected physics processes.


\section{Physics program of the Phase-II Upgrade and motivation for detector upgrades}

Proton-proton collisions with up to 14\,TeV at higher luminosity are foreseen at the HL-LHC.
This brings challenges for the experiments but also prospects for the physics program as described in the following.

\subsection{Challenges for the experiments at the HL-LHC} 

The instantaneous nominal luminosity will increase by a factor five to seven at the HL-LHC compared to the LHC. This will lead to increased particle densities. The increase of the nominal integrated luminosity by a factor of ten will result in increased radiation damage. For the innermost tracking detectors radiation levels of up to  2$\times$10$^{16}$\,1-MeV-equivalent neutrons cm$^{-2}$ or 1\,GRad are expected.
This harsh environment has impact on the detector technologies, the electronics and materials of, for example, cables and glues. All components require a qualification process. 
Moreover, an increase of overlapping proton-proton events (pile-up) from the present average number of proton-proton interactions per bunch crossing $<\mu>\sim$ of about 50 to close to 200 will occur. Additional energy accumulates in the calorimeters, leading to so-called pile-up jets especially in the forward region. 
Hit rates of up to 3\,GHz/cm$^2$ are expected which lead to a higher rate of fake tracks.

\subsection{Physics program}
Both the ATLAS and the CMS experiments foresee to cover a wide physics program spanning over nearly all areas of physics at hadron colliders. Not only the exploration of the electro-weak Standard Model and top physics but also the Higgs Boson program is a major component of the physics program. Rare signatures like Vector-Boson-Scattering and measurements including the Higgs couplings and Higgs self-coupling are covered. Moreover, measurements of QCD processes, constraints of PDF uncertainties with LHC data and constraints on the CKM matrix by measurements in flavour physics are being prepared, for example. 
The HL-LHC will offer an increased data set which reduces uncertainties both statistically and experimentally. Many details can be found in the Refs.~\cite{smhelhc,higgshelhc,bsmhelhc,briefbook}.

\subsection{Motivations for detector upgrades}

A precise measurement of the reconstructed physics objects such as jets, leptons, photons, light-quarks and missing transverse energy is essential to achieve the physics goals. A few specific cases are mentioned as examples in the following.
The efficient tracking with small fake rates requires highly granular tracking detectors which can withstand the radiation doses and have a low material budget. The measurement of the missing transverse energy improves with a high coverage including an acceptance in the forward region of the experiments. Events with high multiplicity and highly boosted jets require also an improved granularity and position resolution of the detector layers.
The physics program including studies of resonances in top pairs, W-, Z-, Higgs and Di-bosons needs the reconstruction of leptons and b-quarks in boosted topologies and a good lepton momentum resolution at high transverse momentum (p$_{T}$).
These cases cannot be covered with the currently installed detectors. Additionally, some sub-detectors cannot withstand the radiation levels and rates expected at HL-LHC, leading to the need to replace or upgrade them.

\section{Detector upgrades}
The ATLAS and the CMS experiments foresee to upgrade their trigger and DAQ systems~\cite{cmsL1TDR,cmsdaq,atlastdaq}.
Moreover, both experiments will replace their tracking detectors with detectors with an extended coverage to $|\eta| < $4~\cite{atlasstripTDR,atlaspixelTDR,cmstrackerTDR}
and include timing detectors~\cite{atlashgtdTDR,cmstimTDR}. The upgrades of these systems will be presented in more detail in the following sections. 
Both calorimeters of the ATLAS experiment will get new front-end electronics~\cite{atlastile,atlaslar}. The front-end electronics of the barrel calorimeter of the CMS experiment will also be renewed~\cite{cmsbarrelcal} while the endcap calorimeters of the CMS experiment will be completely replaced~\cite{cmshgcalTDR}. 
Both experiments will also extend their muon systems with additional units or replacement of electronics~\cite{atlasmuon,cmsmuon}.

\subsection{Upgrade of the trigger and DAQ systems}
\begin{figure}[bp]
\centering
\includegraphics[width=0.48\textwidth]{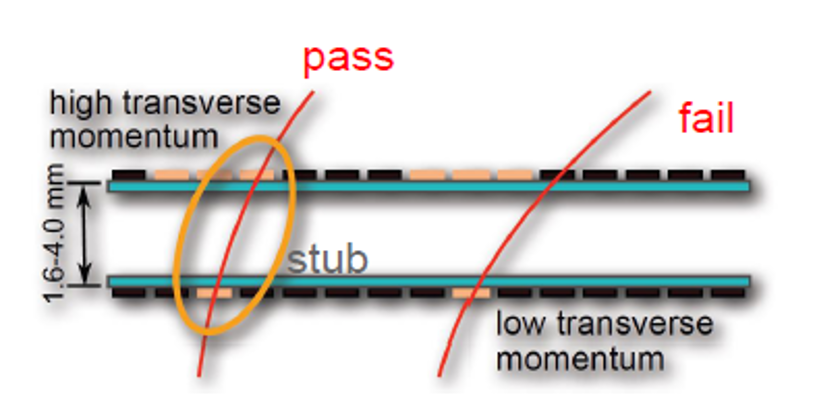} \hfill
\includegraphics[width=0.48\textwidth]{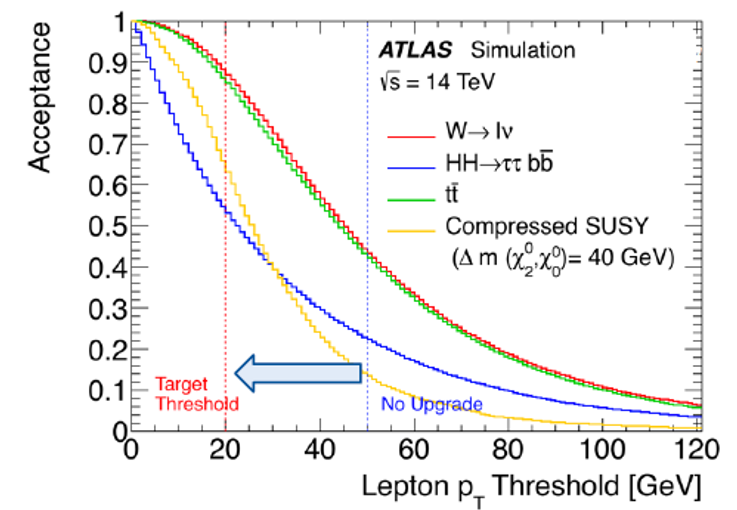} \hfill
\caption{Left: Track trigger of the CMS experiment~\cite{Tricomi}. Right: Acceptance of several physics processes as a function of the lepton p$_{T}$-threshold~\cite{atlastdaq}.  
}
\label{fig:tracktriggercms}
\end{figure}

The upgrade of the trigger and DAQ systems aims to increase the maximum allowable  latencies to enable more complex trigger processing and improve the bandwidth and processing for triggering. 
The ATLAS experiment is expected to process regional tracking of the tracking detector at 1\,MHz, and full granularity global tracking at 100-200\,kHz.
The level-0 triggers will be generated, as in the current trigger scheme, from the calorimeter and muon systems however with an increased latency of 10\,$\mu$s. This is achieved by a partial replacement of the electronics.
A selection based on software will result in an output data rate of 10\,kHz.
The updated trigger selection of the CMS experiment aims to make a selection as close as possible to that of the offline software.
This is achieved by adding tracks in the hardware trigger at the rate of 40\,MHz on level-0. 
A selection of information from the muon system, the calorimeters and tracking detector similar to the particle-flow selection leads to a reduced rate of 750\,kHz with a latency of 12.5\,$\mu$s.
Further rate reduction is achieved with software leading to an output rate of 7.5\,kHz.
The use of tracking information is possible with a dedicated module concept of the tracking detector. 
Modules with two sensors are assembled with a certain spacing and read out with fast electronics.
A schematic view can be seen in Figure~\ref{fig:tracktriggercms}.
Correlations of tracks are calculated on module level and so-called stubs formed. They are sent out if consistent with a track with p$_{T}>$\,2\,GeV. 
By this method the trigger rate can be reduced by about a factor 10~\cite{cmsL1TDR}. 

The upgraded detectors lead to an increase of the acceptance for many physics processes. The Figure~\ref{fig:tracktriggercms} displays the acceptance of several physics processes as a function of the lepton p$_{T}$-threshold~\cite{atlastdaq}.  


\subsection{Upgrade of the tracking detectors and their performance}

The entire ATLAS and CMS tracking systems, which are the sub-detectors closest to the interaction point, will be replaced during the LHC Phase-II shutdown for the operation at the HL-LHC. The aim is to extend the coverage to up to $|\eta|=3.8/4$.
\subsubsection{The new tracking detectors}
The ATLAS experiment will install an all-silicon detector called ITk (Inner Tracker). The ITk consists of an inner pixel and an outer strip detector. The total surface area of silicon in the new pixel system will measure
about 13\,m$^{2}$. The strip detector will comprise 165\,m$^{2}$ of silicon~\cite{atlasstripTDR,atlaspixelTDR}.
The new tracking detector of the CMS experiment will consist of a pixel system of about 4.9\,m$^{2}$, layers of macro-pixel and strip sensors spanning both areas 25\,m$^{2}$, and a strip detector which consists of 138\,m$^{2}$ of silicon modules. All layers except the pixel sensors are foreseen to contribute to the stubs~\cite{cmstrackerTDR}. 
The systems have in common that they foresee to use n-in-p silicon sensors.
More details of the silicon sensors are listed in Table~\ref{table:tracker}.
The systems will deploy fast data transmission with low power giga-bit data transmission, serial powering in the pixel detectors and DC-DC converters for the strips systems.
The cooling will be achieved with CO$_{\mathrm{2}}$ which allows for thinner pipes than mono-phase cooling. 
Carbon structures maintain the mechanical stability resulting in low material budget.
\begin{table}[b]
  \caption{Overview of sensor parameters of the tracking detectors~\cite{atlasstripTDR,atlaspixelTDR,cmstrackerTDR}.} 
  \centering
  \begin{tabular}{|c|c|c|}
    \hline  
    Parameter & CMS & ATLAS  \\ 
    \hline 
    Strip pitch ($\mu$m) & 90-100 & 70-85  \\ 
    Strip length (cm) & 2.5-5 & 2.5-8  \\ 
    Strip thickness ($\mu$m) & 300 & 300  \\ 
    Pixel size ($\mu$m$^{2}$) & 25x100, & 50x50 (planar sensors layers 1-4), \\ 
     &  1.5 mm macro-pixels & 3D-sensors layer 0 in   \\ 
     &  & rings 50x50, in flat 25x100   \\ 
    Pixel thickness ($\mu$m) & $\leq$ 150 &  $\leq$ 150 \\ 
    \hline
  \end{tabular}
\label{table:tracker}
\end{table}

\subsubsection{The tracking performance}

High tracking efficiencies and low fake rates are expected with the new tracking detectors.
The reconstruction efficiency is simulated to be above 90\% in the central region and above 80\% in the forward regions for both new tracking detectors~\cite{atlastrkperf,cmstrackerTDR}.
\begin{figure}[tbp]
\centering
\includegraphics[width=0.56\textwidth]{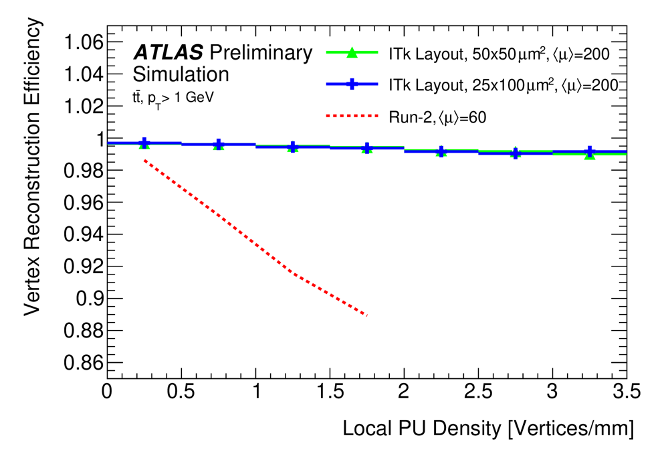} 
\includegraphics[width=0.4\textwidth]{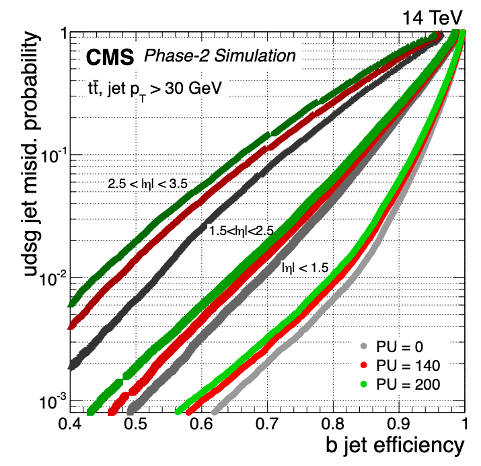} 
\caption{Left: The primary vertex reconstruction efficiency for $t\overline{t}$~\cite{atlastrkperf}. Right: jet mis-identification probability as a function of the b-jet identification efficiency for different regions of $|\eta|$~\cite{cmstrackerTDR}.}
\label{fig:tracking}
\end{figure}
The vertex reconstruction efficiency for $t\overline{t}$ events shows that the tracker can cope with a high number of pile-up events as shown in the left figure of Figure~\ref{fig:tracking}.
The right panel of Figure~\ref{fig:tracking} shows the jet mis-identification probability as a function of the b-jet identification efficiency for different regions of $|\eta|$. The b-tagging performance and light-jet rejection is also robust to pile-up as indicated with differently colored bands corresponding to three levels of pile-up. These quantities are important to discriminate between events from Vector Boson Fusion and $t\overline{t}$ in the forward regions.
Crucial parameters for further pile-up suppression are also expected to improve in comparison to the current tracking detectors. For example, the longitudinal impact parameter (z$_{\mathrm{0}}$ resolution) will be improved in the ATLAS experiment due to reduced material and better resolution of the strip tracker compared to the current transition radiation tracker.

\subsection{Use of timing information}

The high granularity timing detector (HGTD) will be a new detector in the forward region of the ATLAS experiment which is based on low-gain avalanche silicon sensors (LGAD). Disks will cover the range of $ 2.4 < \eta < 4 $ with 2(3) hits per track for radii of $R>30$\,cm ($R<30$\,cm). They will consist of about 6.4\,m$^{2}$ of sensors~\cite{atlashgtdTDR}.
A challenge is the radiation hardness of the components and partial replacement is foreseen after every 1000\,$\mathrm{fb}^{-1}$. The timing resolution per track of 30-50\,ps will allow pile-up effects to be mitigated and to improve the  track-to-vertex resolution. In this way the jet- and e/$\gamma$-reconstruction and e/$\gamma$-isolation will be improved.
The CMS experiment foresees a MIP timing detector with a barrel timing layer from LYSO and silicon photomultipliers and an endcap timing layer with LGAD sensors.
This detector will be placed in the space between the tracker and the calorimeters  and with a timing resolution of 30\,ps will provide the same benefits of the HGTD in ATLAS~\cite{cmstimTDR}.
 
\subsection{New endcap calorimeter of the CMS experiment}

\begin{figure}[h]
\centering
\includegraphics[width=0.98\textwidth]{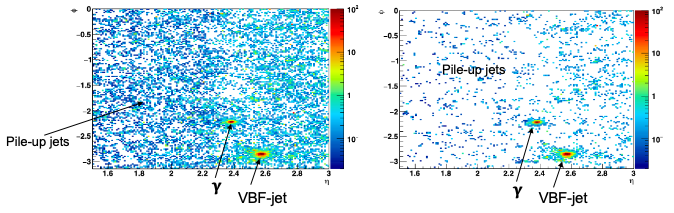}
\caption{Left: Event without timing requirement (including average pile-up of 200)~\cite{cmshgcalTDR}. Right: Event after removal of all hits above 12\,fC threshold with $ \mid \delta t \mid  >$9\,ps~\cite{cmshgcalTDR}.
}
\label{fig:hgcal}
\end{figure}

The upgrade of the tracker of the CMS experiment will be complemented with a new calorimeter with extended coverage to 
$|\eta| < 3 $, high granularity and hence high energy resolution. The so-called high granularity calorimeter~\cite{cmshgcalTDR} is driven by activities of the CALICE collaboration~\cite{calice}.
It is a sampling calorimeter with silicon sensors and scintillators with each technology deployed according to the expected radiation dose. It will consist of an electromagnetic calorimeter built from silicon, Cu/CuW/Pb absorbers with 28 layers and a X$_{\mathrm{0}}$ and $\lambda$ of about 1.7, followed by a hadronic calorimeter with 22 layers of silicon and scintillators with steel absorbers. The total $\lambda$ (from electromagnetic and hadronic calorimeters) is about 9.5.
The active detector area will be 600\,m$^{2}$ of hexagonal 8\,inch sensors and 520\,m$^{2}$ scintillator tiles
connected to SiPMs.
This detector will allow 3D-shower topologies to be reconstructed and give a time resolution of $\sim$30\,ps for particles with a transverse momentum above a few GeV. The topology of a particle shower can be seen in Figure~\ref{fig:hgcal}, which shows reconstructed events of the Vector Boson initiated process for
 H$\rightarrow \gamma \gamma$. The left figure visualizes an event without timing requirement (including average pile-up of 200). The right figure displays an event after removal of all hits above 12\,fC threshold with $ \mid \delta t \mid  >$9\,ps. The improved particle identification is visible. The upgraded detector will also result in a higher energy resolution and pile-up rejection.

\section{Physics prospects}

The upgraded detectors clearly impact and enhance the physics reach at the HL-LHC. A few examples of the wealth of expected results will be illustrated in the following subsections.
\subsection{Physics of the Higgs Boson}
High precision measurements are anticipated for the Higgs couplings which are highly sensitive to physics beyond the standard model. The precision will be significantly improved with the dataset of the HL-LHC. 
The observation of H$ \rightarrow \mu  \mu$ is expected to be possible with a significance of more than nine standard deviations and the uncertainty on the Higgs production cross section times the branching ratio to dimuons normalised by the Standard Model prediction is expected to be around 13\%~\cite{Hmumu}. 
The relative precision on Higgs coupling modifiers $ \kappa$ with $ \kappa_{\nu}  \leq$1 is expected to improve by a factor of two from the LHC to the HL-LHC. It will be constrained on a 2-7\%-level~\cite{briefbook}.
This will also give access to the direct coupling of the Higgs boson to the top-quark with a precision of about 4\% for $\kappa_{t}$. 

The production of Di-Higgs gives a direct handle on the Higgs self coupling. A combination of all channels yields a sensitivity of 4$ \sigma$ using the full dataset of the HL-LHC~\cite{cmsHH,atlasHH}.

\subsection{Vector Boson Scattering and the mass of the W-Boson}

Vector boson scattering is expected to be observable at the HL-LHC  in the leptonic signatures. The expected precision for scattering of WZ-Bosons is about 6\% and 8-10\% for ZZ-Boson scattering, depending on the theoretical uncertainties. 
A precision of below 10\% is calculated for scattering of WW-Bosons with a sensitivity of 2$\sigma$ to W$_{\mathrm{L}}$W$_{\mathrm{L}}$. 
 The challenge in this process is the extraction of the longitudinal scattering component to test unitarity.
Despite major improvements with improved forward tracking and jet-tagging capabilities, the analysis of WW-Bosons will be systematically limited~\cite{cms1805,cms1814,cms1829,cms1838,cms1811}.

The precision measurement of the mass of the W-Boson would be possible at the HL-LHC with a short run (about  200\,$\mathrm{pb}^{-1}$) of low luminosity ($<\mu>=$\,2). Uncertainties down to 9.3\,MeV on the mass are expected~\cite{genprospects}. The improvement is caused by stronger constraints on the PDFs with the extended forward tracking.

\section{Summary}

The HL-LHC will increase the physics reach of the ATLAS and CMS experiments but it also poses challenges to the detectors. Increased integrated luminosity, high pile-up, particle production rates and radiation doses require the upgrade of several sub-detectors in ATLAS and CMS experiments.
These are namely, upgraded electronics for triggering, leading to increased thresholds of triggers and latencies, new electronics for the muon systems, additional muon chambers and new calorimeter readouts in both experiments.
New tracking detectors will be installed with a higher granularity and increased coverage, especially in the forward region. 
The forward tracking will be extended with timing layers. The CMS experiment builds in addition new high granularity calorimeter endcaps. 
These upgrades give access to several rate-limited processes like vector boson scattering and Di-Higgs production. The measurement precision will be improved for several processes, e.g. coupling parameters of the Higgs boson from 15\% at the LHC to a few percent with the full HL-LHC dataset. The reach will also be enhanced for physics beyond the Standard Model. It is additionally essential to further reduce experimental and theoretical uncertainties to fully exploit the rich physics program of the HL-LHC.



\addcontentsline{toc}{chapter}{References}


\begin{thebibliography}{99}

\bibitem{lhc}
L. Evans and P. Bryant (editors), \emph{LHC machine}, \emph{JINST} {\bf \textbf{3} S08001} (2008).

\bibitem{hllhc}
\emph{The High Luminosity LHC project},
Online: https://hilumilhc.web.cern.ch/ (2021).

\bibitem{atlas}
ATLAS Collaboration, \emph{The ATLAS Experiment at the CERN Large Hadron Collider}, \emph{JINST} {\bf \textbf{3} S08003} (2008).

\bibitem{cms}
CMS Collaboration, \emph{The CMS Experiment at the CERN LHC}, \emph{JINST} {\bf \textbf{3} S08004} (2008).

\bibitem{smhelhc}
P. Azzi et al., \emph{Standard Model Physics at the HL-LHC and HE-LHC},
{\tt arXiv:1902.04070 [hep-ph]} (2019).

\bibitem{higgshelhc}
M. Cepeda et al., \emph{Higgs Physics at the HL-LHC and HE-LHC},
{\tt arXiv:1902.00134 [hep-ph]} (2019).

\bibitem{bsmhelhc}
X. Cid Vidal et al., \emph{Beyond the Standard Model Physics at the HL-LHC and HE-LHC}, {\tt arXiv:1812.07831 [hep-ph]} (2019).

\bibitem{briefbook}
European Strategy for Particle Physics Preparatory Group, \emph{Physics Briefing Book}, {\tt arXiv:1910.11775 [hep-ex]} (2020).

\bibitem{cmsL1TDR}
CMS Collaboration, \emph{The Phase-2 Upgrade of the CMS Level-1 Trigger},
CERN-LHCC-2020-004, {\bf CMS-TDR-021}, https://cds.cern.ch/record/2714892.

%
\bibitem{cmsdaq}
CMS Collaboration, \emph{The Phase-2 Upgrade of the CMS DAQ Interim Technical Design Report}, CERN-LHCC-2017-014, {\bf CMS-TDR-018}, https://cds.cern.ch/record/2283193.

\bibitem{atlastdaq}
ATLAS Collaboration, \emph{Technical Design Report for the Phase-II Upgrade of the ATLAS TDAQ System}, CERN-LHCC-2017-020, {\bf ATLAS-TDR-029}, https://cds.cern.ch/record/2285584.

\bibitem{atlasstripTDR}
ATLAS Collaboration, \emph{Technical Design Report for the ATLAS Inner Tracker Strip Detector},
CERN-LHCC-2017-005, {\bf ATLAS-TDR-025}, https://cds.cern.ch/record/2257755.

\bibitem{atlaspixelTDR}
ATLAS Collaboration, \emph{Technical Design Report for the ATLAS Inner Tracker Pixel Detector}, CERN-LHCC-2017-021, {\bf ATLAS-TDR-030}, https://cds.cern.ch/record/2285585.

\bibitem{cmstrackerTDR}
CMS Collaboration, \emph{The Phase-2 Upgrade of the CMS Tracker},
CERN-LHCC-2017-009, {\bf CMS-TDR-014}, https://cds.cern.ch/record/2272264.

\bibitem{atlashgtdTDR}
ATLAS Collaboration, \emph{A High-Granularity Timing Detector for the ATLAS Phase-II Upgrade}, CERN-LHCC-2020-007, {\bf ATLAS-TDR-031}, https://cds.cern.ch/record/2719855.

\bibitem{cmstimTDR}
CMS Collaboration, \emph{A MIP Timing Detector for the CMS Phase-2 Upgrade},
CERN-LHCC-2019-003, {\bf CMS-TDR-020}, https://cds.cern.ch/record/2667167.

\bibitem{atlastile}
ATLAS Collaboration, \emph{Technical Design Report for the Phase-II Upgrade of the ATLAS Tile Calorimeter}, CERN-LHCC-2017-019, {\bf ATLAS-TDR-028}, https://cds.cern.ch/record/2285583.

\bibitem{atlaslar}
ATLAS Collaboration, \emph{ATLAS Liquid Argon Calorimeter Phase-II Upgrade: Technical Design Report}, CERN-LHCC-2017-018, {\bf ATLAS-TDR-027}, https://cds.cern.ch/record/2285582.

\bibitem{cmsbarrelcal}
CMS Collaboration, \emph{The Phase-2 Upgrade of the CMS Barrel Calorimeters},
CERN-LHCC-2017-011, {\bf CMS-TDR-015}, https://cds.cern.ch/record/2283187.

\bibitem{cmshgcalTDR}
CMS Collaboration, \emph{The Phase-2 Upgrade of the CMS Endcap Calorimeter},
CERN-LHCC-2017-023, {\bf CMS-TDR-019}, https://cds.cern.ch/record/2293646.

\bibitem{atlasmuon}
ATLAS Collaboration, \emph{Technical Design Report for the Phase-II Upgrade of the ATLAS Muon Spectrometer}, CERN-LHCC-2017-007, {\bf ATLAS-TDR-026}, https://cds.cern.ch/record/2285580.

\bibitem{cmsmuon}
CMS Collaboration, \emph{The Phase-2 Upgrade of the CMS Muon Detectors},
CERN-LHCC-2017-012, {\bf CMS-TDR-016}, https://cds.cern.ch/record/2283189.

\bibitem{Tricomi}
A. Tricomi, \emph{Upgrade of the {CMS} tracker}, \emph{JINST} {\bf \textbf{9} C03041} (2014).

\bibitem{atlastrkperf}
ATLAS Collaboration, \emph{Expected Tracking Performance of the ATLAS Inner Tracker at the HL-LHC}, {\bf ATL-PHYS-PUB-2019-014}, https://cds.cern.ch/record/2669540.

\bibitem{calice}
\emph{The CALICE Collaboration}, Online: https://twiki.cern.ch/twiki/bin/view/CALICE (2021).

\bibitem{Hmumu}
ATLAS Collaboration, \emph{Prospects for the measurement of the rare Higgs boson decay $H\to\mu\mu$ with 3000 fb$^{-1}$ of $pp$ collisions collected at $\sqrt{s} = 14$ TeV by the ATLAS experiment}, {\bf ATL-PHYS-PUB-2018-006}, https://cds.cern.ch/record/2319741.

\bibitem{cmsHH}
CMS Collaboration, \emph{Prospects for HH measurements at the HL-LHC}, {\bf CMS-PAS-FTR-18-019}, https://cds.cern.ch/record/2652549.

\bibitem{atlasHH}
ATLAS Collaboration, \emph{Measurement prospects of the pair production and self-coupling of the Higgs boson with the ATLAS experiment at the HL-LHC}, {\bf ATL-PHYS-PUB-2018-053}, https://cds.cern.ch/record/2652727.

\bibitem{cms1805}
CMS Collaboration, \emph{Study of W$^\pm$W$^\pm$ production via vector boson scattering at the HL-LHC with the upgraded CMS detector}, {\bf CMS-PAS-FTR-18-005}, https://cds.cern.ch/record/2646870.

\bibitem{cms1814}
CMS Collaboration, \emph{Vector Boson Scattering prospective studies in the ZZ fully leptonic decay channel for the High-Luminosity and High-Energy LHC upgrades}, {\bf CMS-PAS-FTR-18-014}, https://cds.cern.ch/record/2650915.

\bibitem{cms1829}
CMS Collaboration, \emph{Search for excited leptons in llgamma final states in proton-proton collisions at the HL-LHC}, {\bf CMS-PAS-FTR-18-029}, https://cds.cern.ch/record/2652017/.

\bibitem{cms1838}
CMS Collaboration, \emph{Prospects for the measurement of electroweak and polarized WZ to 3l$\nu$ production cross sections at the High-Luminosity LHC}, {\bf CMS-PAS-FTR-18-038}, https://cds.cern.ch/record/2650774.

\bibitem{cms1811}
CMS Collaboration, \emph{Sensitivity projections for Higgs boson properties measurements at the HL-LHC}, {\bf CMS-PAS-FTR-18-011}, https://cds.cern.ch/record/2647699.

\bibitem{genprospects}
A. Dainese, M. Mangano, A.Meyer, A. Nisati, G. Salam and M. Vesterinen, \emph{Report on the Physics at the HL-LHC, and Perspectives for the HE-LHC}, {\bf CERN-2019-007}, https://cds.cern.ch/record/2703572.




\end{thebibliography}
\end{document}